\documentclass{elsart}
\usepackage{graphicx}
\usepackage{dcolumn}
\usepackage{bm}
\usepackage[latin2]{inputenc}
\usepackage[english]{babel}
\usepackage{amssymb}
\journal{Physica D}
\begin{document}

\begin{frontmatter}

\title{HebbNets: Dynamic network with Hebbian learning rule}
\author{G. Szirtes}
\author{Zs. Palotai}
\author{A. L{\H o}rincz\corauthref{cor1}}
\ead{lorincz@inf.elte.hu}
\address{Department of Information Systems, E\"otv\"os Lor\'and
University, P\'azm\'any P\'eter s\'et\'any 1/C, Budapest
Hungary,H-1117} \corauth[cor1]{corresponding author Address:
Department of Information Systems, E\"otv\"os Lor\'and University,
P\'azm\'any P\'eter s\'et\'any 1/C, Budapest Hungary, H-1117
P:36-1-2090555 ext 8483}

\date{\today}

\begin{abstract}
It has been demonstrated that one of the most striking features of
the nervous system, the so called 'plasticity' (i.e high
adaptability at different structural levels) is primarily based on
Hebbian learning which is a collection of slightly different
mechanisms that modify the synaptic connections between neurons.
The changes depend on neural activity and assign a special dynamic
behavior to the neural networks. From a structural point of view,
it is an open question what network structures may emerge in such
dynamic structures under 'sustained' conditions when input to the
system is only noise. In this paper we present and study the
`HebbNets', networks with random noise input, in which structural
changes are exclusively governed by neurobiologically inspired
Hebbian learning rules. We show that Hebbian learning is able to
develop a broad range of network structures, including scale-free
small-world networks.
\end{abstract}

\end{frontmatter}

\section{\label{s:intro}Introduction}
In the last few years research on complex interactive systems
(CISs) has become one of the most fascinating areas. One generally
applied way to describe such systems is based on graphs with nodes
(vertices) and (directed) edges, representing constituents of the
system and their interactions. Classification of CISs is grounded
on their structural and dynamic network properties. Similar
network structures may be found in many different fields spanning
from social connection systems to biochemical processes
\cite{Watts98Collective,kleinberg98authoritative,albert99diameter,barabasi99emergence,barabasi00scalefree,marchioria00harmony,Latora01Efficient,Bohland01Efficient,FerrerCancho01Optimization,albert02statistical}.
Due to the recognition of many common characteristics in both
natural and human (artificial) CISs, several general models have
been designed to describe the emergence of such structures, e.g.,
by random restructuring of the links among a finite number of
`nodes' ~\cite{Watts98Collective} or by 'preferential attachment'
~\cite{albert99diameter,barabasi99emergence}, or by optimizing the
link structure of finite systems
~\cite{FerrerCancho01Optimization}.

Probably the most complex network is inside us: the most exciting
properties of our brain have a lot to do with the special
connection system among its units. It is widely accepted that
activity correlation between the computing units (i.e. different
forms of the so called Hebbian learning mechanism) plays a
fundamental role in forming the complex neural structures and
maintaining its intrinsic plasticity. Its essence is that the
connection strength between the communicating units is modified
according to the simultaneous activity correlation of the signal
sender and receiver.

It is worth noting that the concept of Hebbian learning has
undergone revolutionary changes in the last few years. The
original suggestion of Hebb \cite{Hebb49Organization} has been
modified by recent findings
\cite{markram97regulation,magee97synaptically,bell97synaptic}. For
a review, see, e.g., \cite{abbott00synaptic}. A unifying
description is called spike-time dependent synaptic plasticity
(STDP) and it allows different time shift patterns between the
units' activities.

\section{\label{s:model}Description of HebbNet}

In this letter we examine what network structures may emerge in a
simplistic neural system by applying \textbf{pure} Hebbian
dynamics without any special additional constraints. This neuronal
network model will be referred as to \textit{HebbNet}. We assume
that the network is \textit{sustained} by inputs with no
spatio-temporal structure; the input is random noise. Our models
consist of $N$ number of simplified integrate-and-fire like
`neurons' or nodes. The dynamics of the internal activity is
written as
\begin{equation}
 \frac{\Delta a_i}{\Delta t} =\sum_{j} w_{ij}a_j^s+x_i^{(ext)},
 \label{e:int_and_fire}
\end{equation}
for $i=1,2,\ldots ,N$. (N was 200 in our simulations.) Variable
$x^{(ext)} \in {(0,1)}^N$ denotes the randomly generated input
from the environment, $a_i$ is the internal activity of neuron
$i$, $w_{ij}$ is $ij^{th}$ element of matrix $\mathbf{W}$, i.e.,
the connection strength from neuron $j$ to neuron $i$. If $\Delta
t = 1$ then we have a discrete-time network and each parameter has
a time index, or if $\Delta t$ is infinitesimally small then
Eq.~\ref{e:int_and_fire} becomes a set of coupled differential
equations. The neuron $j$ outputs a spike (neuron $j$ fires) when
$a_j$ exceeds a certain level, the threshold parameter $\theta$.
Spiking means that the output of the neuron $a_j^s$ (superscript
$s$ stands for 'spiking') is set to 1. Otherwise, $a_j^s=0$.
Amount of excitation received by neuron $i$ from neuron $j$ is
$w_{ij} a_j^s$ when neuron $j$ fires. After firing, $a_j$ is set
to zero at the next time step. For continuous case $a_j$ is set to
zero after a very small time interval. Equation
\ref{e:int_and_fire} describes the simplest form of
`integrate--and--fire' network models which is still plausible
from a neurobiological point of view. No temporal integration
occurs for the discrete case provided that the left hand side of
Eq.~\ref{e:int_and_fire} is replaced by $a^{+}_i$ where
superscript $+$ denotes time shifting. In this limiting case, and
if the threshold is high enough, `binary neurons' emerge.  This
model resembles the original model of McCullough and Pitts
\cite{McCullough43Logical}.

\begin{figure}
  \centering
  \includegraphics[width=6cm]{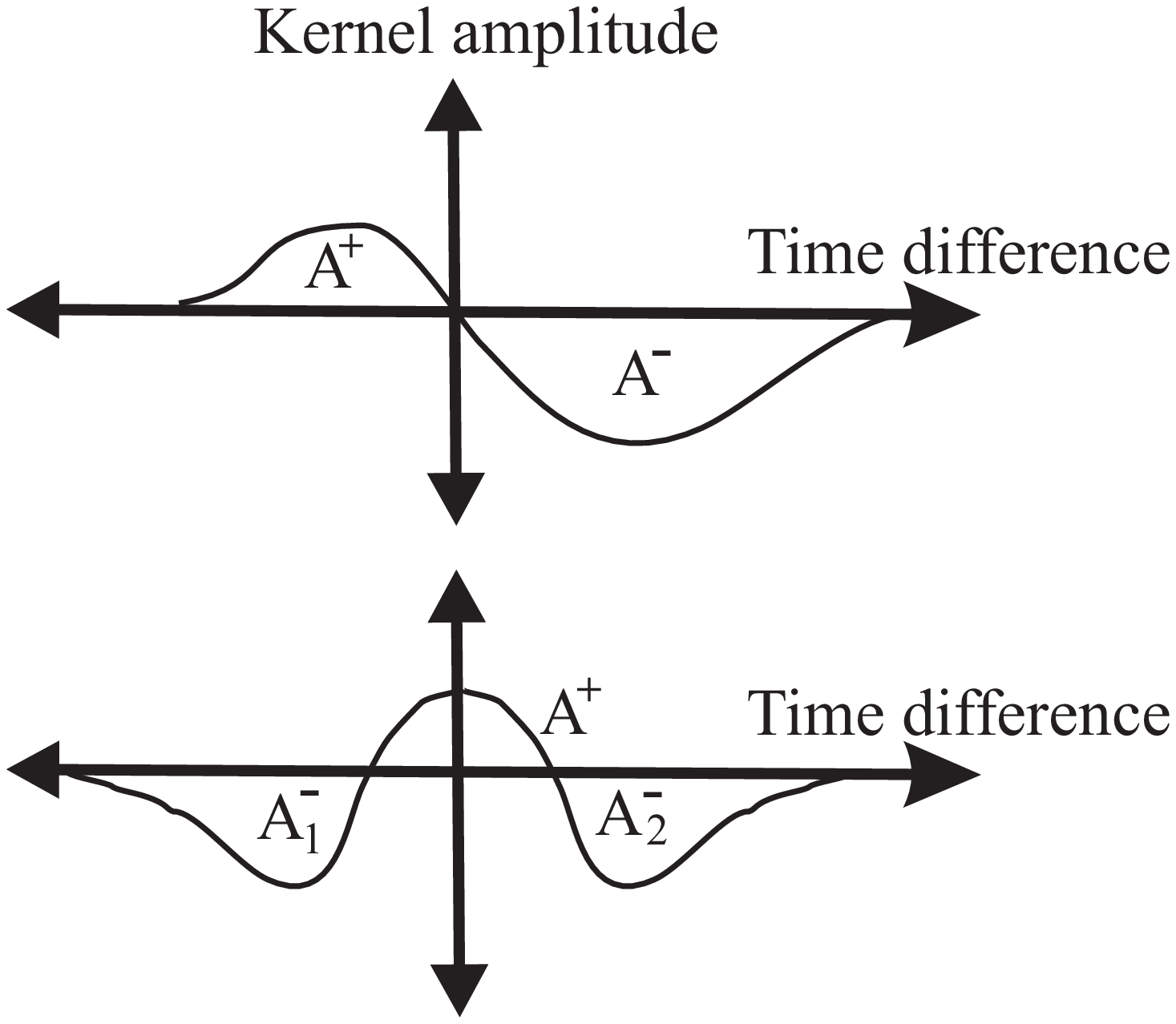}
  \caption{\textbf{Kernel functions}
  \newline Two temporal kernels as a function of time difference between
  spiking time of neuron $i$ and $j$ ($t_i-t_j$). Relevant
  parameter of the shape for noise-sustained systems is
  the ratio ($r_{A^{+}/A^{-}}$) of the areas/sums of positive  and negative parts/components of the kernel,
  $A^{+}$ and $A^{-}$, respectively ($r_{A^{+}/A^{-}}=A^{+}/A^{-}$).
  }\label{f:kernelf}
\end{figure}

We examined the effect of local activity threshold and global
activity constraint (selection of a given percent of nodes with
the highest activity). The former one is more realistic
biologically, while the latter one is more convenient: in this way
the ratio of active units is always known and fixed. For these two
cases, computer simulations showed negligible differences.
Synaptic strengths were modified as follows:
\begin{equation}
 \frac{\Delta w_{ij}}{\Delta t}
=\sum_{(t_i,t_j)}K(t_j-t_i)a_i^{t_i,s}a_j^{t_j,s},
 \label{e:pot}
\end{equation}
where $K$ is a kernel function which defines the influence of the
temporal activity correlation on synaptic efficacy and $\Delta
w_{ij} / \Delta t$ may be taken over discrete or over
infinitesimally small time intervals. Possible examples are
depicted in Fig.~\ref{f:kernelf}. The kernel is a function of the
time differences. When the input is made of noise, as in our
studies, only the ratio of the positive (strengthening) and the
negative (weakening) parts of the kernel function should count.
This is the result of the lack of temporal correlations in the
input. Temporal grouping and reshaping of the kernel would not
modify our results as long as the said ratio is kept constant. In
turn, our results concern both types of kernels depicted in
Fig.~\ref{f:kernelf}.

In the first place, we have been interested in the emerging local
and global connectivity structure of $\mathbf{W}$. Instead of
using global structural property ($L$, characteristic path length
which is the average number of edges on the shortest paths) and
the clustering coefficient ($C$) proposed by Watts and Strogatz
\cite{Watts98Collective} we applied the so called connectivity
length measure based on the concept of \textsl{network efficiency}
\cite{Latora01Efficient}. This single measure is more appropriate
for weighted networks \cite{marchioria00harmony}, equally well
applicable for describing global and local properties and offers a
unified theoretical background to characterize our system.
According to the definition
~\cite{marchioria00harmony,Bohland01Efficient}, local efficiency
between nodes $i$ and $j$ in a weighted network with connectivity
matrix $\mathbf{W}$ is $\epsilon_{ij}= 1/d_{ij}$, where $d_{ij}=
\min_{n,\,k_1, \ldots k_n} \left( 1/w_{ij}, 1/w_{ik_1} + \ldots +
1/w_{k_{n-1}k_n} + 1/w_{k_n j}\right)$ ($k_m \in (1,2, \ldots N)$
for every $1 \leq m < N-1 $ and $1 < n \leq N$). For graphs with
connection strengths of values 0 or 1, $d_{ij}$ corresponds to the
\textit{shortest distance} between nodes $i$ and $j$. The average
of these values ($E[d_{ij}] =\frac{1}{N(N-1)}\sum_{i\neq
j}\epsilon_{ij}$) characterizes the efficiency of the whole
network. The local harmonic mean \textit{distance} for node $i$ is
defined as
\begin{equation}
 D_h(i) =
\frac{n^{(i)}}{\sum_{j : w_{ij} > 0}\epsilon_{ij}},
\end{equation}
where $n^{(i)}$ is the number of neurons around neuron $i$ with
$w_{ij}>0$. In terms of efficiency, this inverse of this value
describes how good the local communication is amongst the first
neighbors of node $i$ with node $i$ removed. It is a measure of
the fault tolerance of the system. The mean \textit{global
distance} in the network is defined by the following quantity:
\begin{equation}
D_h = \frac{N(N-1)}{\sum_{i,j}\epsilon_{ij}}.
\end{equation}
Global distance provides a measure for the \textit{size} (or the
diameter) of the network, which influences the average time of
information transfer. According to
\cite{marchioria00harmony,Bohland01Efficient} local harmonic mean
distance measure behaves like $1/C$ (inverse of the clustering
coefficient), whereas the global value corresponds to $L$. It can
be shown that L is a good approximation of $D_h$ (or $1/L$ for the
global efficiency) under certain conditions
\cite{Bohland01Efficient}.

\section{Results and Discussion}\label{s:results}

\begin{figure}[h!]
  \centering
  \includegraphics[width=8cm]{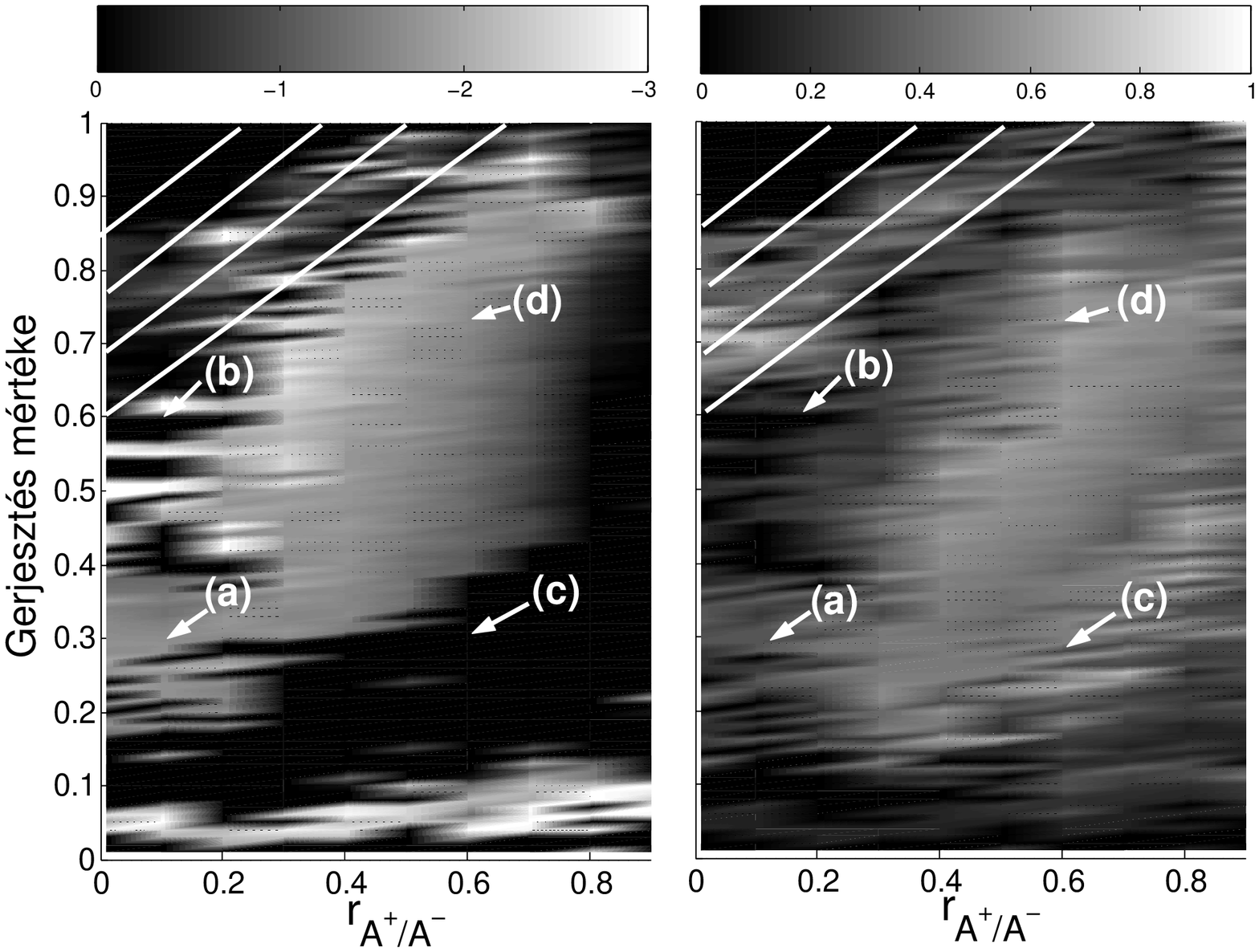}
  \caption{\textbf{Scale-free region with negligible interaction}\newline
  \textbf{Left:} \textit{exponent of the power law}, \textbf{right:} \textit{relative percentage of the power-law domain}
  as a function of $r_{A^+/A^-}$ and $r_{ex}$ (the ratio of excited neurons).
  Contribution of other neurons to the neuronal inputs is negligibly small. Difference between
  binary and integrate-and-fire neurons disappears in this limiting
  case. Results are averaged over 20 runs, all sampled 50 times, $\theta=0.5$. Stripes denote unstable region:
  components of matrix $\mathbf{W}$ may vanish. Log-log plots corresponding to points (a)--(d) are
  shown in Fig.~\ref{f:loglog}. Power-law with negative (positive) exponent: cases (a) and (d) (case (c)).
  Positive exponents are thresholded to zero on the figure. For
  visualization purposes, the data have been interpolated between
  the calculated grid points.
  }\label{f:nofeedback}
\end{figure}

\begin{figure}[h!]
  \centering
  \includegraphics[width=8cm]{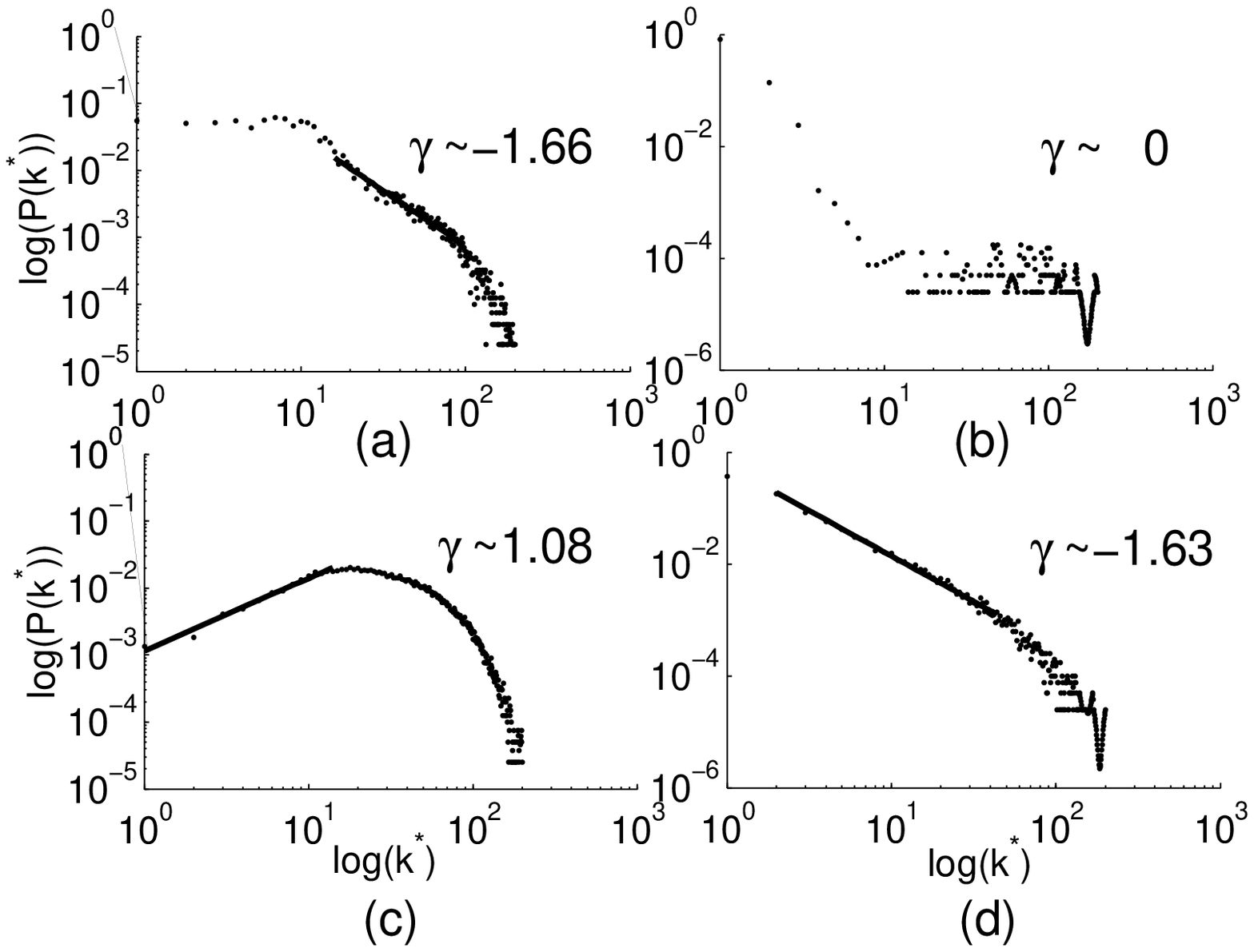}
  \caption{\textbf{Log-log plots for different parameters}\newline
  The four diagrams display typical distributions ($P(k^*)$) for
  parameters shown in
  Fig.~\ref{f:nofeedback} by (a), (b), (c) and (d).
  Cases (a) and (d) are arbitrary examples from the power law region.
} \label{f:loglog}
\end{figure}

These connectivity length measures allowed us to study the
emerging network structures as the function of the following
parameters: (i) the magnitude of the external excitation (defined
by the average percentage of neurons receiving excitation from the
environment and (ii) the strengthening--weakening area ratio of
the kernel, $K$. The binary neuron model was also investigated.
Figures \ref{f:nofeedback} and \ref{f:loglog} summarize our
findings in different parameter regions. The figure displays the
appearance of scale free nets as a function of the excitation
level and $r_{A^+/A^-}$. The length of the scale-free regions was
determined by first plotting the distribution of the sum of the
weights of outgoing connections (averaged over 10000 samples taken
from 20 networks) for every parameter set studied. Results were
depicted on loglog plot. Supposing a power-law distribution
($P(k^*)\approx k^{*-\gamma}e^{-k^*/\xi}$, where $k^*$ denotes the
discretized values of the connection strength), a linear fitting
was made to approximate $\gamma$. The width of the scale-free
region was estimated by the length of the region with power-law
distribution relative to the full length covered on the log scale.
Maximum error of the linear fit was set to $10^{-3}$ STD. That is,
for 100 discretization points, the width of a region spreading an
order of magnitude on the loglog plot is equal to 0.5.


\begin{figure}[!ht]
  \centering
  \includegraphics[width=6cm]{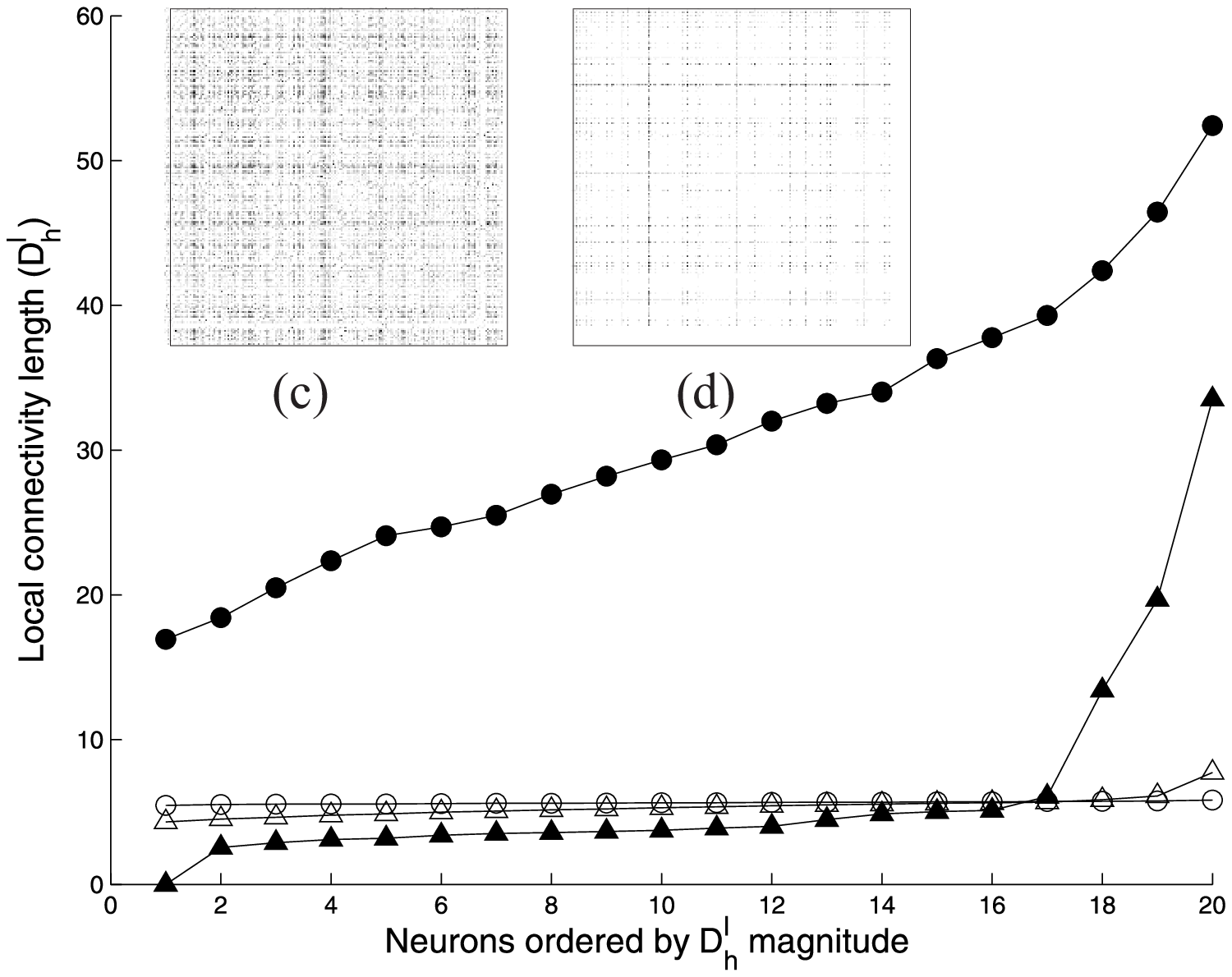}
  \caption{\textbf{Harmonic mean distances}\newline
  Local harmonic mean distances in ascending order are shown.
  For better visualization not all data points are marked and the points are connected with
  a solid line. Lines with upward triangle markers: STDP learning. Lines with circles:
  same but randomly redistributed weights. Line with empty (solid) markers: HebbNet of
  case (c) (case (d)). Global harmonic mean distances for the original and
  for the randomized networks in case (c) of Fig.~\ref{f:nofeedback} (case (d) of Fig.~\ref{f:nofeedback})
  are about the same $D_h\approx D_h^r \approx 5.5$ ($D_h \approx D_h^r \approx 10$).
  The two inlets show the resulting connection matrices.
  }\label{f:Harmdis}
\end{figure}

Fig.~\ref{f:Harmdis} displays the emerging connections of a
HebbNet for two different parameter sets. We compared the
resulting HebbNet structures with a random net, in which the same
weights of the dynamic network have been randomly assigned to
different node pairs. The two inlets show the HebbNet connection
matrices. While inlet (c) belonging to case (c) in
Fig.~\ref{f:nofeedback} resembles a random connection matrix,
inlet (d) belonging to case (d) in Fig.~\ref{f:nofeedback}
represents a sparse structure. (Note that most elements are not
zero, but very small.)

Fig.~\ref{f:Harmdis} highlights clearly the emerging small-world
properties, i.e., small local connectivity values (high clustering
coefficients) for case (d). Although the global connectivity
length was almost the same for all HebbNets and their
corresponding random nets, local distances are much smaller in
case (d). That is, connectivity structure is sparse but
information flow is still fault tolerant and efficient.

The robustness of the network to the external excitation is
illustrated on the next figure.
\begin{figure}[!ht]
  \centering
  \includegraphics[width=6cm]{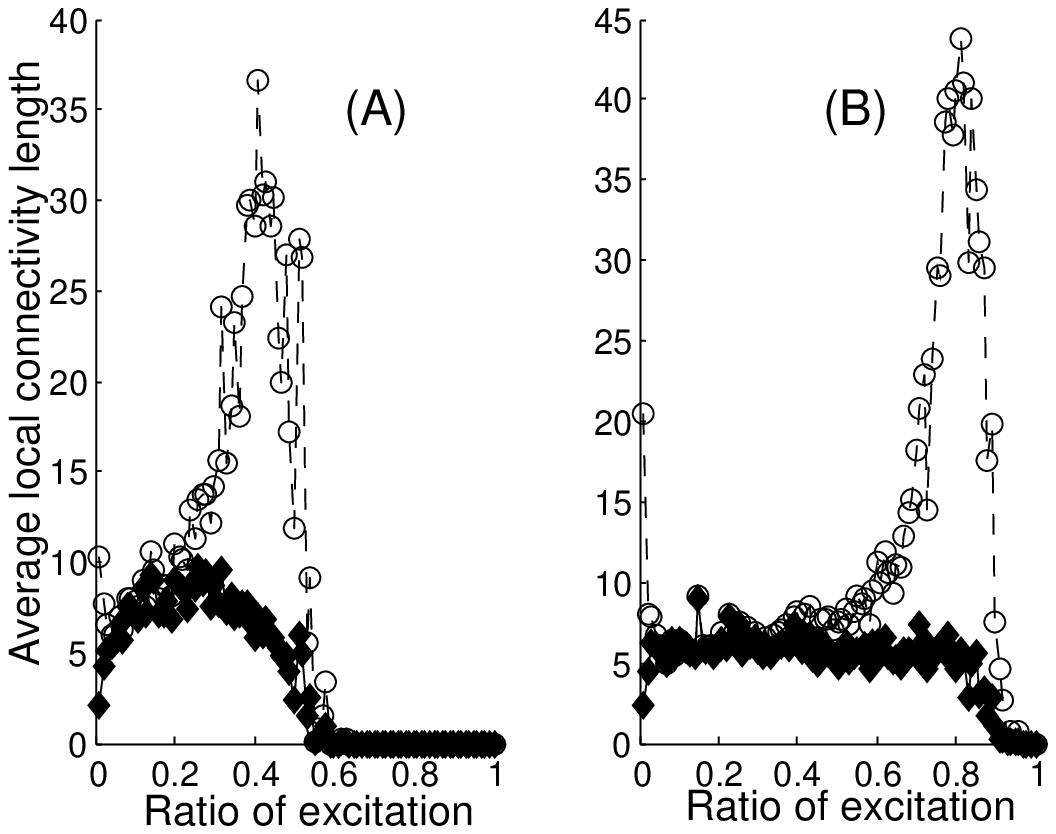}
  \caption{\textbf{Average local distance vs. excitation ratio}\newline
  \newline
  A: $r_{A^+/A^-}=0.1$, B: $r_{A^+/A^-}=0.6$.
  Diamonds: average local distances for the evolving network.
  Circles: average local distances for the corresponding random net.
  }\label{f:robust}
\end{figure}
By increasing the excitation level, the average local connectivity
length of the random net is drastically increasing, whereas the
efficiency of the small-world network does not change too much in
the same region. For the network with parameters $r_{A^+/A^-}=0.1$
(Fig.~\ref{f:robust}(A)), there is a sharp cut-off around
excittion level 0.55, where local distances suddenly drop, due to
the high ratio of excitation. Qualitatively similar behavior can
be seen for $r_{A^+/A^-}=0.6$ (Fig.~\ref{f:robust}(B)), but the
cut-off is around $r_{ex}=0.9$.

\begin{figure}[h!]
  \centering
  \includegraphics[width=8cm]{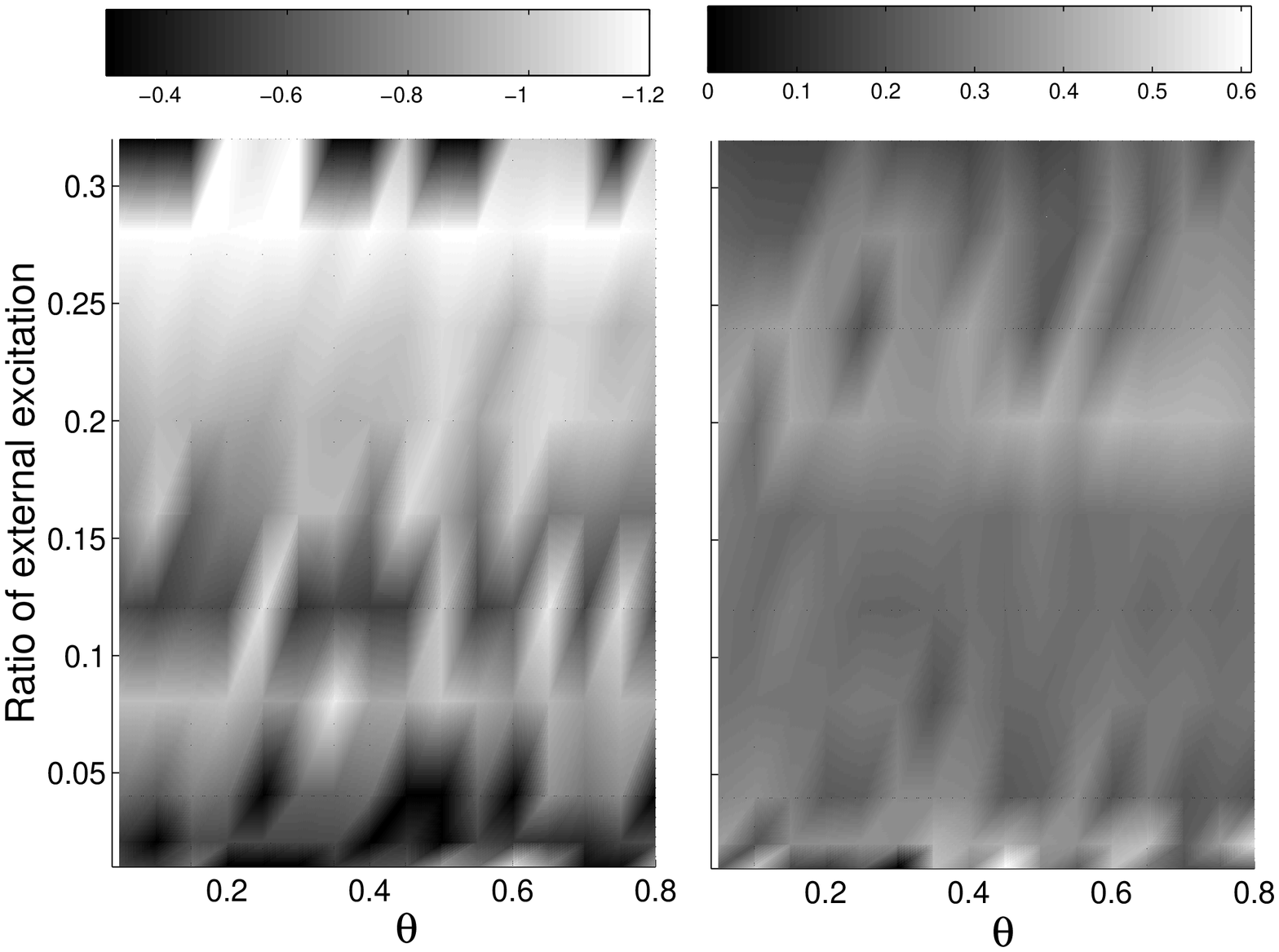}
  \caption{\textbf{Power-law with significant interaction}\newline
  \textbf{Left:} \textit{exponent of the power law},
  \textbf{right:} \textit{relative percentage of the power-law domain}
  as a function of $r_{ex}$ and excitation threshold $\theta$. $r_{A^+/A^-}=0.1$
  Results are averaged over 700 steps. Input from other neurons could exceed
  the external inputs by a factor of 10. The exponent of the power-law
  approximates -1 for broad regions of $\theta$ and $r_{ex}$.
  Outside this region the network may vanish or may start to oscillate. For
  visualization purposes, the data have been interpolated between
  the calculated grid points.
}\label{f:feedback}
\end{figure}

For networks with significant interaction we have experienced a
convergence of the exponent of the power-law distribution to -1.
The width of the scale-free region was relatively broad (see,
Fig.~\ref{f:feedback}).

\section{Summary}
In summary, we have demonstrated that small-world architectures
with scale-free domains may emerge in sustained networks under
STDP Hebbian learning rule without any other specific constraints
on the evolution of the net. Although one always has to remember
that results from simplified models may not carry over to
biophysically realistic networks, we feel that some intriguing
conjectures can be made based on our findings. The role of noise
in the central nervous system
~\cite{Ferster96Neuralnoise,Miller02Neuralnoise} is unclear. The
existence of such `HebbNets' may support the speculative view of
Kandel et al. \cite{Kandel92Adult} that structural development and
learning plasticity in CNS may have a common basis. According to
our results, evolution and plasticity of the networks may be
maintained by noise randomly generated within the CNS. We
conjecture that the sustained nature of noise and the competition
imposed by small $r_{A^+/A^-}$ values are the two relevant
components of plasticity and learning. It might be equally
important that exponents of HebbNets with significant interaction
amongst neurons are similar in a broad range of parameters.

As far as other evolving networks are considered, the profound
implication of our result is that local (Hebbian) learning rules
may be sufficient to form and maintain an efficient network in
terms of information flow. This feature differs from existing
models, such as the model on preferential attachment
\cite{barabasi99emergence}, the global optimization scheme
\cite{FerrerCancho01Optimization}, and also from the original
Watts and Strogatz model \cite{Watts98Collective}.

\section{Acknowledgements}
\vspace{0.5cm}  This work was partially supported by the Hungarian
National Science Foundation, under Grant No. OTKA 32487.

\bibliographystyle{elsart-num}
\bibliography{hebbnet_physicaD}

\end{document}